# High resolution x-ray investigation of periodically poled lithium tantalate crystals with short periodicity

M. Bazzan,[1] C. Sada,[1,a)] N. Argiolas,[1] A. C. Busacca,[2] R. L. Oliveri,[2] S. Stivala,[2] L. Curcio,[2] and S. Riva Sanseverino[2]

[1]*Dipartimento di Fisica "G. Galilei," Università di Padova and CNISM, Via Marzolo 8 35131 Padova, Italy*
[2]*Department of Electrical, Electronic and Telecommunication Engineering, DIEET, University of Palermo, Viale delle Scienze Bldg. 9, 90128 Palermo, Italy*

Domain engineering technology in lithium tantalate is a well studied approach for nonlinear optical applications. However, for several cases of interest, the realization of short period structures ($<2$ $\mu$m) is required, which make their characterization difficult with standard techniques. In this work, we show that high resolution x-ray diffraction is a convenient approach for the characterization of such structures, allowing us to obtain in a nondestructive fashion information such as the average domain period, the domain wall inclination, and the overall structure quality.

## I. INTRODUCTION

Ferroelectric crystals have been used in a number of electronic and optoelectronic applications since their electro-optical, photorefractive, and piezoelectric properties, as well as their quadratic nonlinear response, are quite large and can also be tailored with a simple inversion of ferroelectric domains which can be engineered in both 1D[1,2] and 2D.[3] A challenge in ferroelectric domain poling is the realization of short periods ($<2$ $\mu$m) while preserving good uniformity and repeatability. Surface periodic poling (PP) is a convenient approach to the engineering of ferroelectric structures with periods as short as 400 nm.[4,5] Breaking the micron-period barrier for periodical bulk domain patterning is also very desirable for several applications such as tunable cavity mirrors, first order backward second harmonic generation, and pulse shaping by backward frequency doubling of femtosecond laser.[6–8]

Among the others, lithium tantalate (LT) (congruent type composition) is one of the most promising ferroelectric crystals in the field of nonlinear integrated optics since it exhibits excellent nonlinear optical properties, with high threshold to photorefractive damage, extended transparency in the UV down to 280 nm,[9] and large electro-optic and quadratic responses $d_{33}=15.1$ pm/V at 852 nm.[10–12] As a consequence, the possibility of realizing periodical inversion of the ferroelectric spontaneous polarization in LT crystals (PPLT) is of great interest, as already demonstrated for designing coherent optical sources in the ultraviolet UV.[12]

One of the key point in the PP process is the characterization of the periodically poled structures based on nondestructive methods opposed to the standard commonly used destructive fluoridric acid (HF) attack. Several methods have been proposed, and, for an extensive review, the readers are referred to a specific literature (e.g., Ref. 13). Among the others, we shall mention here, as the most widespread approaches, optical techniques, scanning probe techniques, and structural techniques. Optical techniques are based on the direct observation of domain structures by some kind of optical microscopy. These techniques are difficult to be used for characterization of submicrometric domain patterns since their spatial resolution is not sufficient at this length scale. Moreover, these techniques do not probe the effective reversal of the nonlinear optical coefficients, but optical features which are indirectly connected to change of polarization, such as the refractive index changes induced at the domain walls, and this fact hinders somehow the interpretation of the experimental data. A possible alternative is constituted by scanning probe microscopy, which provides a nanoscale characterization of electrical, mechanical, and optical properties of ferroelectric materials but has the drawback that it allows only a local investigation of the ferroelectric domains, while for Quasi Phase Matching (QPM) applications, a characterization of the "average" properties of the periodic structures would be more desirable. Structural characterizations are generally performed by studying the scattering of x-ray and taking advantage that a reversal of the spontaneous polarization is directly related to a change of the structure factor of the crystalline matrix.[14] This fact is generally used to perform a topographic imaging of the periodically poled structure, but again, the limited spatial resolution of the imaging system forbids the application of this approach to sub micrometric structures. If small sized domain structures are to be probed, it is preferable to use a reciprocal space approach, because in this case the smaller the details to be probed, the easier the detection and the measure of periodic structures, owing to the reciprocal character of the relation linking the direct space to the reciprocal space.[15] In other words, a short period gives rise to well separated features in reciprocal space.

In this work, we exploit the high resolution x-ray diffraction (HR-XRD) technique to perform a reciprocal space analysis of periodically poled LT crystals prepared by elec-

[a)]Author to whom correspondence should be addressed. Electronic mail: cinzia.sada@unipd.it. Tel.: +39-049-8277037. FAX: +39-049-8277003.

tric field poling. While the periodicity of the satellites allows to determine straightforwardly the period of the PP, the intensity and the shape of the satellites are affected by the waveform of the polarization profile (duty cycle, etc.), perfection of the structure, presence of strain, domain walls inclination, and so on, and can therefore be interpreted only if a model of the structure is provided. In this work, we test the method in a critical situation, i.e., with PP structures with a period as large as 1.5 $\mu$m, giving rise to reciprocal space maps whose details are close to the resolution limit in reciprocal space of our apparatus. Thanks to reciprocal space map analysis, the technique also allows in this case to extract important structural information. For structures with smaller periodicities, the technique is even more appropriate and informative.

## II. EXPERIMENTAL

The PPLT samples were diced from 500 $\mu$m thick, Z-cut, congruent LT wafers. We used high voltage pulses applied across the LT thickness in order to achieve surface poling with domains as deep as 40 $\mu$m.[16] A 1.3 $\mu$m-thick photoresist (Shipley S1813) was spin coated on the $-Z$ face of the crystal and a 1.5 $\mu$m period was defined using an amplitude mask to transfer the pattern by UV lithography. The periodic pattern was 7 mm long and 100 $\mu$m wide. After development, the hardness and the adhesion of the photoresist was improved by baking the sample overnight at 90 °C and then hard baking it at 130 °C for 3 h. In this phase, considering that LT has a strong pyroelectric effect, temperature processes need to be performed gradually, in order to avoid the formation of poling dots, mechanical stress, and trapping of charges on the crystal surfaces. During high voltage poling, the photolithographic mask acts as the insulating layer necessary to produce and periodically control the domain inversion. A waveform generator (Agilent 3220A) and a high voltage amplifier (Trek 662) produce the electric field to pole the sample. Electric contact between electrodes and the substrate was ensured by a conductive saline gel mixture. To exceed the LT coercive field (22 kV/mm), we applied single 1.3 kV pulses over a 10 kV bias for appropriate time intervals, the latter adjusted to fulfill the poling condition $Q < 1.5 A P_s$, with $Q$ as the total flowing charge, $A$ as the poled area, and $P_s$ as the spontaneous polarization of the ferroelectric crystal. The flow of charge was opportunely controlled by a home-designed analog current integrator. The waveforms of both the poling voltage and the flowing current were registered with a digital oscilloscope. Using this approach, the inverted domains nucleated from the $-Z$ facet in the region under the electrodes and extended toward the $+Z$ facet. The scanning electron microscopy picture of Fig. 1 shows the PPLT region of a typical sample after selective chemical etching in HF at 70 °C for 15 min.

The x-ray investigations were performed using a Philips MRD diffractometer with a sealed Cu anode source, equipped with a parabolic multilayer mirror for enhanced beam intensity (see Fig. 2). The beam is collimated and monochromatized using a four-bounce 2 2 0 channel-cut Ge monochromator giving a primary beam with a wavelength $\lambda = 0.154\,056$ nm, a spectral purity $\Delta \lambda / \lambda = 2 \times 10^{-5}$ and an angular divergence of 0.0032°. The detector was a proportional counter equipped with a three bounce Ge 2 2 0 analyzer whose acceptance is equal to the divergence of the primary beam. Due to the very large acceptance of the detector on the axial direction, the measurements reported hereafter have to be considered as the projection of the scattered intensity on the plane defined by the goniometers movements (the so-called scattering plane), reducing in this way the three-dimensional reciprocal space to a single plane. In order to avoid artifacts during the measurements due to thermal drifts, the temperature of the measure chamber was stabilized at $25.0 \pm 0.1$ °C. Using a crossed slit arrangement, the primary beam footprint was confined to an area of about $0.1 \times 5$ mm$^2$ and made to impinge upon the area of the sample where the periodically poled structure was present. The symmetrical 0 0 12 reflection of LT was measured in reciprocal space mapping mode.[17] In particular, with reference to Fig. 2, we remind that by defining $\alpha_i$ and $\alpha_f$ as the angles between the incident beam and the diffracted beam, respectively, with the sample surface, $\omega$ as the angle of the goniometer in the sample stage with respect to the primary beam and $2\theta$ as the angle of the detector goniometer with respect to the primary beam, the following relations hold:

$$\omega = \alpha_i, \quad (1)$$

$$2\theta = \alpha_i + \alpha_f, \quad (2)$$

$$Q_x = K(\cos \alpha_i - \cos \alpha_f), \quad (3)$$

$$Q_z = K(\sin \alpha_i + \sin \alpha_f), \quad (4)$$

where $K = 2\pi/\lambda$ is the wavenumber of the incident radiation and $Q_x$ and $Q_z$ are the coordinates in the scattering plane of the reciprocal space point probed by the instrument. By pro-gramming the movements of the goniometers, a given region of the reciprocal space can be investigated by associating to each couple $(Q_x,Q_z)$ the intensity $J(Q)$ of the radiation accepted by the detector, so that a map of the reciprocal space intensity distribution is recorded.

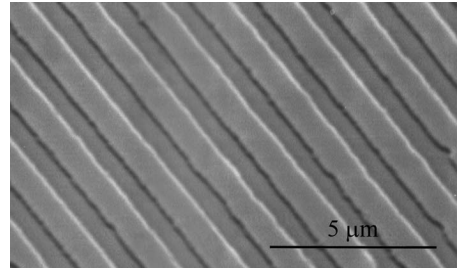

FIG. 1. Scanning electron microscope image of a 1.5 $\mu$m periodic structure on LT as revealed after HF acid attack.

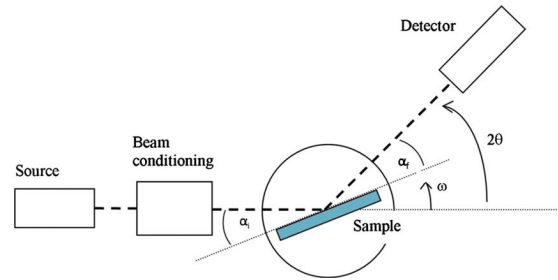

FIG. 2. (Color online) Sketch of the experimental HR-XRD apparatus.J.

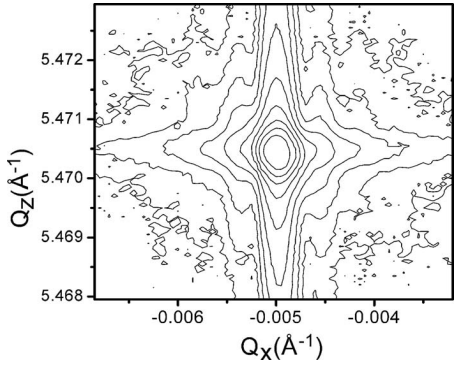

FIG. 3. Reciprocal space map of the 0 0 12 reflection. The iso-intensity contours are plotted on a logarithmic scale with the maximum intensity normalized to one. Each contour represents a four-time increase of the intensity.

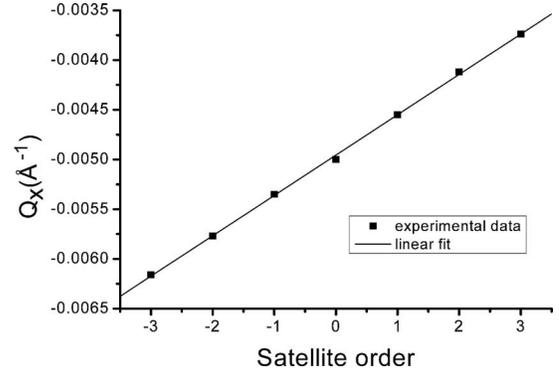

FIG. 5. Plot of the $Q_x$ position of the satellite maxima as a function of the satellite order. The straight line is a linear fit to the experimental data. The error bars, calculated *a posteriori*, are smaller than the experimental point symbols..

## III. RESULTS AND DISCUSSION

The resulting map is shown in Fig. 3. The map is composed of a central rod and a series of closely spaced satellite rods parallel to the $Q_z$ direction. Moreover, a horizontal streak crossing the rods is clearly visible. Since the area illuminated by the x-ray beam is much larger than the area containing the PP structures, the recorded signal contains contributions coming from both the periodic structure and the unpoled substrate, and the former contribution has a considerably lower intensity with respect to the substrate peak. Anyway, it can be evidenced by projecting the map on the $Q_x$ axis that an intensity modulation superimposed to the flanks of the substrate peaks is clearly observable (see Fig. 4). The overall quality of the PP structure is confirmed by the presence of several satellite peaks, which can be observed up to the fourth order. The 0-th order streak corresponds to the "average" lattice of the superstructure and does not show any peak splitting within the experimental resolution of the reciprocal space map, indicating that the average "perpendicular" lattice parameter (namely, the $c$ parameter) of the PP structure is practically indistinguishable with respect to that of the unpoled region, i.e., no mismatch induced by residual stress due to poling treatment is present.

The horizontal streak visible in the map is the diffuse (incoherent) scattering originating from the domain wall regions.[18] This peak has an elliptical shape with the major axis parallel to $Q_x$ axis and then to the sample surface, which allows us to conclude that the domain walls are perpendicular to the sample surface within experimental uncertainty ($\sim 0.1°$).

As the micron-sized period of the PP structures gives rise to closely spaced structures in reciprocal space, the experimental resolution of our instrument did not allow us to resolve the satellite maxima. However, the diffuse scattering streak indicates that the grating wavevectors must contain no $Q_z$ components, otherwise the domain walls could not be aligned perpendicular to the sample surface. As a consequence, the satellite maxima must be aligned parallel to the $Q_x$ direction, on a horizontal line crossing the central maximum of the map.

Taking this fact into account, the lateral positions of the satellite maxima can be estimated by checking the intersection of the horizontal axis of the diffuse scattering ellipse with the vertical satellite rods; the resulting positions are shown as arrows in Fig. 4. By plotting their position as a function of the satellite order (Fig. 5), we get a straight line whose slope is given by the modulus $q=2\pi/\Lambda$ of the fundamental wavevector of the PP grating, where $\Lambda$ is the period of the PP. From this, we get $\Lambda=1.547\pm 0.023$ $\mu$m, in good agreement with the period of the amplitude mask used to fabricate the domain structures and the period of domains structures as revealed after etching in HF.

## IV. CONCLUSIONS

We have shown here that the HR-XRD technique in reciprocal space mapping mode can be successfully employed for the nondestructive characterization of periodically poled structures in LT. Owing to the reciprocal character of this analysis, this kind of approach is particularly suited for PP structures with small period. In this work, we studied a sample with a period of about 1.5 $\mu$m, which is close to the resolution limit of the technique. Even in this extreme case, informations such as the overall quality of the structure, the domain wall inclination, and the structure period can be obtained. In samples with smaller periodicities or shallower domains, the technique would work even better, as the satellite structure in that case can be completely resolved. In principle, determination of the strain state of the superlattice, the domain depth, fluctuations in superlattice periods, and so on can be extracted using methodologies analogs to those routinely employed for the x-ray characterization of other superlattice structures such as multiquantum wells or surface gratings.[17,19]

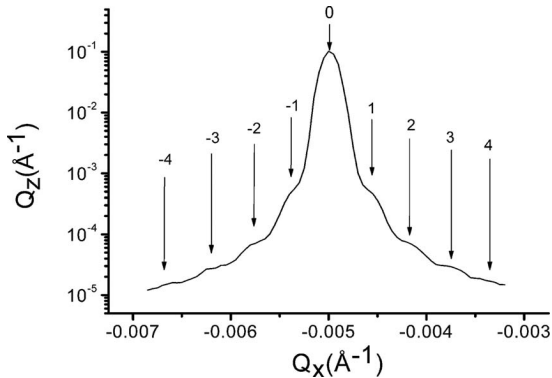

FIG. 4. Projection of reciprocal space map of Fig. 1 onto the $Q_x$ axis. The satellite structure is evidenced by arrows indicating the satellite order.

## ACKNOWLEDGMENTS

The work was partially funded by the Italian Ministry for Scientific Research (MiUR) through Grant No. 2007CT355. We are grateful to Dott. A. C. Cino for his support.